\documentclass[osajnl,twocolumn,showpacs,superscriptaddress,10pt]{revtex4-1} 
\usepackage{amsmath,amssymb,graphicx}
\usepackage{geometry} 
\usepackage{color}
\begin{document}

\title{Lensfree Spectral Light-field Fusion Microscopy for Contrast- and Resolution-enhanced Imaging of Biological Specimens}

\author{Farnoud Kazemzadeh}\email{Corresponding author: fkazemzadeh@uwaterloo.ca}
\affiliation{Department of Systems Design Engineering, University of Waterloo, 200 University Ave. West, Waterloo, Ontario, N2L 3G1}
\author{Chao Jin}
\affiliation{Department of Civil and Environmental Engineering, University of Waterloo, 200 University Ave. West, Waterloo, Ontario, N2L 3G1}
\author{Sara Molladavoodi}
\author{Yu Mei}
\affiliation{Department of Systems Design Engineering, University of Waterloo, 200 University Ave. West, Waterloo, Ontario, N2L 3G1}
\author{Monica B. Emelko}
\affiliation{Department of Civil and Environmental Engineering, University of Waterloo, 200 University Ave. West, Waterloo, Ontario, N2L 3G1}
\author{Maud B. Gorbet}
\author{Alexander Wong}
\affiliation{Department of Systems Design Engineering, University of Waterloo, 200 University Ave. West, Waterloo, Ontario, N2L 3G1}

\begin{abstract}

A lensfree spectral light-field fusion microscopy (LSLFM) system is presented for enabling contrast- and resolution-enhanced imaging of biological specimens.  LSLFM consists of a pulsed multispectral lensfree microscope for capturing interferometric light-field encodings at various wavelengths, and Bayesian-based fusion to reconstruct a fused object light-field from the encodings.  By fusing unique object detail information captured at different wavelengths, LSLFM can achieve improved resolution, contrast, and signal-to-noise ratio (SNR) over a single-channel lensfree microscopy system.  A five-channel LSLFM system was developed and quantitatively evaluated to validate the design. Experimental results demonstrated that the LSLFM system provided SNR improvements of 6-12 dB, as well as a six-fold improvement in the dispersion index (DI), over that achieved using a single-channel, resolution-enhancing lensfree deconvolution microscopy system or its multi-wavelength counterpart. Furthermore, the LSLFM system achieved an increase in numerical aperture (NA) of $\sim$16\% over a single-channel resolution-enhancing lensfree deconvolution microscopy system at the highest-resolution wavelength used in the study. Samples of \emph{Staurastrum paradoxum}, a waterborne algae, and human corneal epithelial cells were imaged using the system to illustrate its potential for enhanced imaging of biological specimens.
\end{abstract}


\maketitle 

Optical microscopy remains an essential imaging technique for many fields of research and technology.  Various developments are underway in this field, with three important areas of development being: i) increasing the field-of-view (FOV), ii) reducing instrument complexity, and iii) increasing imaging resolution.  An emerging microscopy modality that has shown considerable promise at addressing the first two areas is lensfree microscopy, where the concept of holographic imaging is harnessed to capture interferometric light-field encodings without the need of lenses, from which complex object light-fields can be reconstructed from the encodings.  By forgoing the need for lenses, the instrument complexity of lensfree microscopy systems can be greatly reduced while the FOV can be greatly increased.  Given these attractive properties, there has been widespread interest in adaptation of such devices in fields such as biology, histology, particle distribution and motion, and water quality assessment~\cite{Su13, Frentz10, Wong15, Mudanyali10, Greenbaum14, Sheng06, Kiss13}.  One important caveat of lensfree microscopy systems is that imaging resolution is limited by the pixel pitch of the sensor array, making it difficult to rival traditional light microscopy systems in terms of imaging resolution.  To address this issue, a number of lensfree microscopy solutions based on synthetic aperture imaging and illumination source scanning and angling have been proposed~\cite{Mico06, Mico10, Granero10, Leon08, Isikman12, Bishra11, Greenbaum13}.  However, such solutions require elaborate mechanical scanning equipment and more complex reconstruction algorithms that greatly increases the complexity and size of such systems.  As such, an alternative mechanism for improving imaging resolution and contrast in lensfree microscopy without the need for elaborate mechanical scanning equipment is highly desired.  One such approach is the resolution-enhancing lensfree deconvolution microscopy~\cite{Greenbaum13}, which is widely accepted to be state-of-the-art in improving contrast and resolution beyond the pixel pitch of the detector without the need for mechanical illumination scanning.

One area that is not well-explored but holds significant promise for lensfree microscopy is the utilization of different wavelengths of light to capture different detailed information about the sample being imaged.  Recently, a few studies have proposed the use of three wavelengths in the visible band (red, green, and blue) as illumination sources for lensfree microscopy, with the primary focus of replicating images captured by traditional optical microscopy~\cite{Isikman12, Kiss13, Isikman10, Greenbaum13a}. Some recent works have also explored the use of different wavelengths to improve phase retrieval~\cite{Noom1,Noom2}.  However, the study of how detailed information captured using different wavelengths can be leveraged to improve resolution and contrast in lensfree microscopy has not been well-explored.

\begin{figure}[!t]
	\centering
    \includegraphics[width=0.7\linewidth]{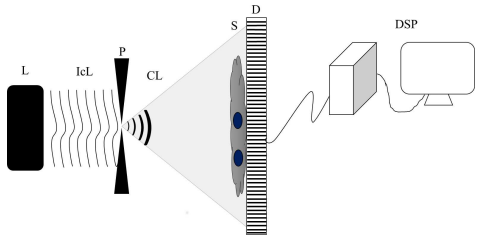}
        \vspace{-0.1in}
	\caption{Experimental pulsed multispectral lensfree microscopy setup for lensfree spectral light-field fusion microscopy.  LEDs at five different wavelengths are contained in the light source box (L) and pulsed to sequentially emit partially-coherent light (IcL).  This light encounters a 200 $\mu$m aperture (P) and becomes spatially coherent (CL), as a result of passing through the aperture, it is then imposed onto the sample (S). The interferometric light-field encodings at each wavelength are then generated on and recorded by detector (D), and sent to the digital signal processing (DSP) unit to perform Bayesian-based fusion to reconstruct the fused complex object light-field.}
	\label{fig1}
\vspace{-0.25in}
\end{figure}

In this study, we aim to introduce a new lensfree spectral light-field fusion microscopy (LSLFM) system for improving imaging resolution and contrast in lensfree microscopy by leveraging unique detailed information associated with differing diffraction behaviour at different wavelengths.  In LSLFM, a pulsed multispectral lensfree microscope first captures interferometric light-field encodings at different wavelengths.  Based on the captured encodings and intrinsic system properties, Bayesian-based fusion is then performed to reconstruct a fused complex object light-field with a higher resolution and contrast to improve sample detail visibility beyond what can be achieved at individual wavelengths.  The LSLFM system does not require any mechanical illumination scanning equipment and as such is especially attractive for applications outside of a laboratory environment where large FOV, high SNR and NA, and excellent image quality are paramount yet instrument complexity and size must be minimized.

For this study, a five-channel pulsed multispectral lensfree microscope was developed (see Fig.~\ref{fig1} for experimental configuration) to examine the efficacy of LSLFM for improved resolution, contrast, and SNR.  A total of five partially-coherent light emitting diodes (LEDs) were used as the pulsed multispectral light source with central wavelengths ranging from the visible band to the near-infrared (NIR) band at $\lambda_{1}=465$ nm, $\lambda_{2}=525$ nm, $\lambda_{3}=591$ nm, $\lambda_{4}=639$ nm, and $\lambda_{5}=870$ nm, with the spectral bandwidth being $\pm$ 30 nm.  The LEDs in the pulsed light source were programmed to pulse using an Arduino mini controller board, with the duration of the pulsations set to maximize the signal observed on the detector while avoiding pixel saturation.  The detector readout was synchronized with the LED pulsations, pending the appropriate exposure time, to facilitate for rapid and seamless acquisitions of interferometric light-field encodings at the five different wavelengths. The shortest exposure time was 750 ms for $\lambda_{2}=525$ nm and the longest was 1 s for $\lambda_{5}=870$ nm. These exposure times are directly related to the quantum efficiency of the detector, which has highest efficiency in the visible green range and the lowest efficiency in the NIR range, and to the power of the LEDs used in the pulsed multispectral light source.  Note that this high exposure time is a major limitation of the present microscope configuration due to the light source and detector used and not due to the LSLFM methodology, and improving both the light source and the detector will allow for much shorter exposure times for imaging dynamic events.

The samples were placed on a \#1 microscope cover slip with a thickness of $\sim$ 145 $\mu$m and placed directly on the detector. Interferometric light-field encodings of the samples were captured by the detector at the five different wavelengths using a 1600$\times$1200 pixel CMOS sensor array with a pixel pitch of 4.5 $\mu$m.  The FOV of the pulsed multispectral lensfree microscope is determined by the active sensor size and is $\sim$ 35 mm$^2$.

\begin{figure}[!t]
	\centering
    \includegraphics[width=1\linewidth]{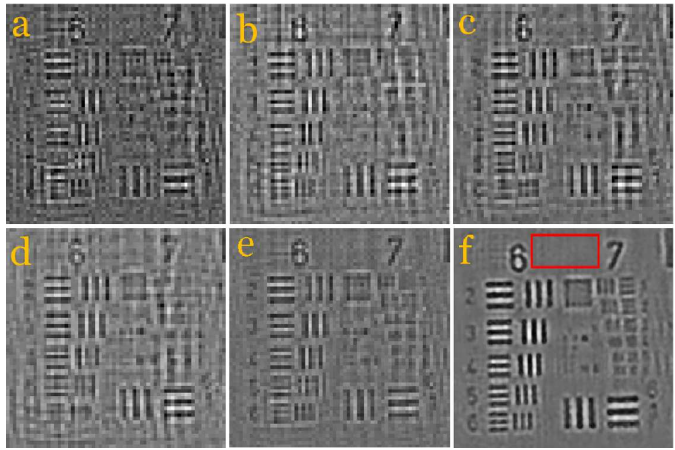}
    \vspace{-0.27in}
	\caption{A region of the USAF resolution target captured using resolution-enhancing LDM system using $\lambda_{1}=465$ nm (a), $\lambda_{2}=525$ nm (b), $\lambda_{3}=591$ nm (c), and $\lambda_{4}=639$ nm (d).  The same region captured using MWLDM system (e), and proposed LSLFM system (f). The SNR and the DI were assessed in a region shown by the red box.}
	\label{fig2}
\vspace{-0.25in}
\end{figure}

Given the set of captured interferometric light-field encodings, denoted by $G=\{g_{\lambda}| \lambda \in \Lambda\}$ where $\Lambda$ denotes the set of wavelengths $\Lambda=\{\lambda_1,\lambda_2,\ldots,\lambda_n\}$, we introduce a Bayesian-based fusion procedure in the DSP unit for reconstructing a fused object light-field $f_{\bigoplus}$ based on the unique diffraction behaviour at different wavelengths as captured in $G$, along with intrinsic system properties. The fusion procedure, which incorporates the statistical models first introduced in~\cite{Wong15} but for the fundamentally different problem of reconstructing an object light-field from multiple interferometric light-field encodings, can be described as follows.  The fused object light-field $f_{\bigoplus}$ can be defined as a weighted integral of a set of object light-fields at different wavelengths (denoted by $F=\{f_{\lambda} | \lambda \in \Lambda\}$):
\begin{equation}
	{f_{\bigoplus}} = \int_{\lambda \in \Lambda} w_{\lambda}f_{\lambda},
\label{super}
\end{equation}
\noindent where $w_{\lambda}$ is the integral weight of $f_{\lambda}$. Let the set of object light-fields at different wavelengths $F$ and the set of captured encodings $G$ be modeled as probability distributions.  The goal is to determine the weighted integral of the set of most probable object light-fields $F$ given $G$, based on \emph{a priori} information related to $F$, \emph{a priori} information about the aberration transfer functions at each wavelength ($H_{a,\lambda}$), and the diffraction transfer functions at each wavelength ($H_{d,z,\lambda}$).  This can be formulated as the following weighted integral optimization problem:
\begin{equation}
	\hat{f}_{\bigoplus} = \int_{\lambda \in \Lambda} w_{\lambda}{\rm argmax}_{f_{\lambda}}~p\left(g_{\lambda} | f_{\lambda}\right)p(f_{\lambda}),
\label{MAP}
\end{equation}
\noindent where $p\left(g_{\lambda} | f_{\lambda}\right)$ is the likelihood of captured encoding $g_{\lambda}$ given the object light-field $f_{\lambda}$ and $p(f_{\lambda})$ is the prior of $f_{\lambda}$.  Based on quantum photon emission statistics, $p\left(g_{\lambda} | f_{\lambda}\right)$ can be expressed by:
{\scriptsize
\begin{equation}
	p\left(g_{\lambda} | f_{\lambda}\right) = \prod_{s \in S} \frac{\left(\mathfrak{F^{-1}}\left\{\frac{H_{a,\lambda}}{H_{d,z,\lambda}}\mathfrak{F}\left\{f_{\lambda,s}\right\}\right\}\right)^{{g_{\lambda,s}}}e^{-\left(\mathfrak{F^{-1}}\left\{\frac{H_{a,\lambda}}{H_{d,z,\lambda}}\mathfrak{F}\left\{f_{\lambda,s}\right\}\right\}\right)}}{{g_{\lambda,s}}!}
\label{likelihood}
\end{equation}
}
\noindent where $\mathfrak{F}$ and $\mathfrak{F^{-1}}$ denotes the forward and inverse Fourier transform, respectively, and $S$ denotes a set of locations in the sensor array, with $s \in S$ being a specific location in the sensor array.  For the prior, we model $f_{\lambda}$ as a nonstationary process with a nonstationary expectation ${E}(f_{\lambda,s})$ and a variance $\tau^2$~\cite{Wong15a}:
\begin{equation}
	p\left(f_{\lambda}\right) = \prod_{s \in S} e^{-\frac{\left(f_{\lambda,s}-{E}(f_{\lambda,s})\right)^2}{2 \tau^2}}.
\label{prior}
\end{equation}

The weighted integral optimization problem posed in Eq.~\ref{MAP} is solved using the iterative optimization method described in~\cite{Wong15a}, with the weights $W=\{w_{\lambda} | \lambda \in \Lambda\}$ determined from the loading obtained via factor analysis~\cite{Spearman04} to account for correlation structure across wavelengths.

\begin{figure}[!t]
	\centering
    \includegraphics[width=1\linewidth]{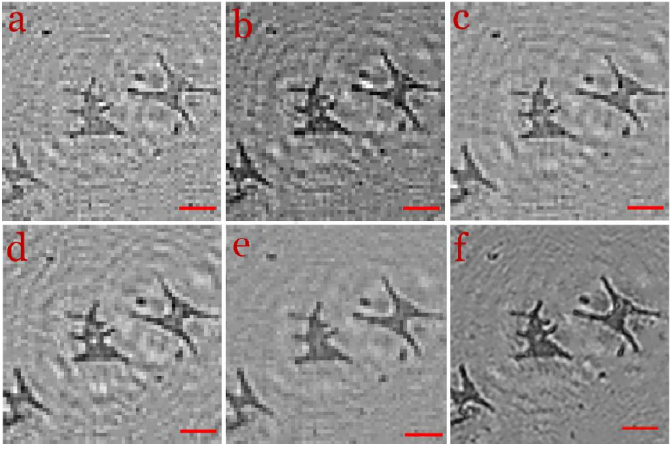}
            \vspace{-0.25in}
	\caption{Selected region of the \emph{Staurastrum paradoxum} sample captured with LDM system using $\lambda_{1}=465$ nm (a), $\lambda_{2}=525$ nm (b), $\lambda_{3}=591$ nm (c), and $\lambda_{4}=639$ nm (d).  The same region captured using MWLDM system (e) and proposed LSLFM system (f).  The scale bar denotes 40 $\mu$m.}
	\label{fig3}
\vspace{-0.25in}
\end{figure}

The efficacy of the proposed LSLFM system was validated using the 1951 USAF resolution target, as well as samples of \emph{Staurastrum paradoxum} (SP), a waterborne algae, and human corneal epithelial cells (HCECs).  For comparison purposes, a single-channel resolution-enhancing lensfree deconvolution microscopy (denoted here as LDM) system using the five different wavelengths was also realized using the same instrumentation setup as the LSLFM system, with the exception of the light pulsations and the DSP unit which are unique to LSLFM.  The deconvolution for the LDM system is achieved via the Maximum-likelihood method described in~\cite{Greenbaum13}, which is widely accepted to be state-of-the-art for improving contrast and resolution beyond the pixel pitch of the detector without the need for mechanical illumination scanning.  The Hyugens numerical diffraction transfer function and the same characterized aberration transfer function were used for both the LDM system and the LSLFM system for consistency.  Furthermore, the iterative optimization methods used in both the LDM and LSLFM system are both performed for 35 iterations for convergence, as suggested by~\cite{Greenbaum13}.  In addition, a multi-wavelength variant of the LDM system, which we will refer to as MWLDM, is realized by combining the complex object light-fields produced across multiple wavelengths using LDM via multi-wavelength averaging.

Fig.~\ref{fig2} shows a zoomed-in region of the 1951 USAF resolution target captured with the LDM system at four of the five captured wavelengths (Fig.~\ref{fig2}(a-d)).  The same region captured using the MWLDM system and the LSLFM system is shown in Fig.~\ref{fig2}e and Fig.~\ref{fig2}f, respectively.  It can be observed that, compared to the results obtained by the LDM system using different wavelengths, the results produced using the LSLFM system exhibits noticeable improvements in contrast, resolution, as well as noticeable reductions in imaging artifacts.  While the MWLDM system exhibits strong improvements in contrast and reduction in imaging artifacts compared to LDM, the LSLFM system is able to provide noticeable improvements in resolution.

The imaging resolution of the LDM system at different wavelengths, the MWLDM system, and the LSLFM system were then determined based on the smallest element in which the vertical lines or the horizontal lines are resolved.  It was determined that the highest NA achieved by the LDM system is $\sim$0.0703 computed at $\lambda_{c}=465$ nm, as the vertical lines of Group 7 Element 1 are resolved.  In comparison, the LSLFM system achieves an NA of $\sim$0.085, as the horizontal lines of Group 7 Element 2 are resolved, which is an improvement of $\sim$16\% over LDM, which is the state-of-the-art in improving resolution without mechanical scanning.  Furthermore, the MWLDM system achieves an NA of $\sim$0.0703, as with LDM, and therefore the LSLFM system achieves an improvement of $\sim$16\% over MWLDM, which also illustrates that the resolution gains of LSLFM are not due to noise reduction and contrast enhancement from multi-wavelength averaging.

The red box shown in Fig.~\ref{fig2}f is the region where the SNR was assessed. The evaluated SNRs in dB are 15.03, 16.42, 17.97, 20.75 for the LDM system using $\lambda_{1}$, $\lambda_{2}$, $\lambda_{3}$, and $\lambda_{4}$, respectively. The MWLDM system achieves an SNR of 21.22 for the same region where as the LSLFM system achieved an SNR of 27.56 dB, resulting in an SNR gain of $>$ 12 and 6 dB over the LDM and the MWLDM systems, respectively. Dispersion Index (DI), described in~\cite{Bianco14}, was assessed for the same region to evaluate noise level, with lower DI indicating lower noise levels. The DIs achieved were 0.0115, 0.0109, 0.007, 0.0043 for the LDM system using $\lambda_{1}$, $\lambda_{2}$, $\lambda_{3}$, and $\lambda_{4}$, respectively.  The DI measured for MWLDM result is 0.0032. In comparison, the LSLFM system achieved a DI of 0.0007, which is six times lower than the best DI achieved using the LDM system and more than four times lower than the MWLDM system.

\begin{figure}[!t]
	\centering
    \includegraphics[width=0.7\linewidth]{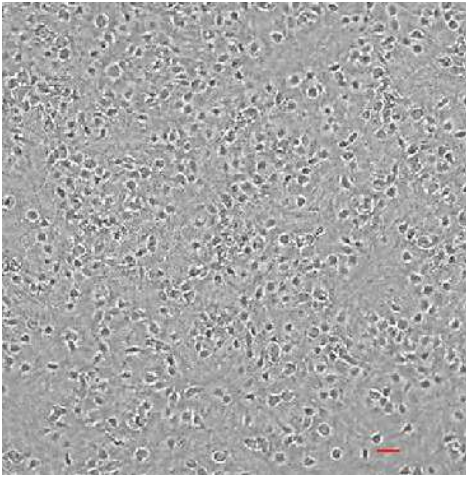}
            \vspace{-0.15in}
	\caption{A selected region of the human corneal epithelial cells captured using LSLFM system with the equivalent FOV of a 40X light microscope. The scale bar denotes 20 $\mu$m.}
	\label{fig4}
\vspace{-0.25in}
\end{figure}

To validate the functionality and merit of the LSLFM system, two biological specimens were imaged using the device. An aliquot laboratory pure culture of SP, a common surface water algae, and a culture of immortalized HCECs were selected as targets for image acquisition (shown in Fig.~\ref{fig3} and Fig.~\ref{fig4}).

\emph{Staurastrum paradoxum} is an algae species belonging to the Desmidiaceae family of green algae, commonly found in Canadian freshwater environments. A pure culture of this algae was obtained from the Canadian Phycological Culture Collection at the University of Waterloo. The culture was propagated in an environmental growth chamber at 23$^{\circ}$C in pre-sterilized BG-11 medium~\cite{Rippka79} to which an aqueous solution of vitamins B12, biotin and thiamine was aseptically added.  The culture flasks were kept under cool-white fluorescent lamps (300 - 400 $\mu$E m$^{-2}$ s$^{-1}$ ) for one week to allow for initial growth and then transferred to lower illumination (200 - 300 $\mu$E m$^{-2}$ s$^{-1}$) for slower growth and storage.

The HCECs were maintained in keratinocyte medium, supplemented with keratinocyte growth supplement and Penicillin/Streptomycin at 37$^{\circ}$C, 5\% CO$_2$, and 95\% humidity. The cell culture medium was replaced every 2-3 days. The HCECs were cultured on 22$\times$22 mm$^2$ \# 1 coverslip one day before the imaging to allow for cells to adhere and spread on the coverslips. To prepare samples for microscopy, cells were fixed with 2.5\% glutaraldehyde solution for two hours and subsequently were washed with phosphate buffer saline solution three times.

Fig.~\ref{fig3} shows a selected region of the SP sample in the same order as presented in Fig.~\ref{fig2}. As compared to the results obtained using the LDM system (Fig.~\ref{fig3}(a-d)) and MWLDM (Fig.~\ref{fig3}e), the results using the LSLFM system (Fig.~\ref{fig3}f) has a noticeably higher contrast and resolution while the majority of imaging artifacts are suppressed.  Many species of algae are used as surrogates and indicators of other algae or contaminants that may be present in water and can be harmful to human and ecosystem health.  The LSLFM system developed herein has the potential to more efficiently and economically provide critical information for waterborne biological contaminant detection, enumeration, and identification.

An image of the HCEC is shown in Fig.~\ref{fig4}. This figure demonstrates a select region, a fraction, of the total FOV captured with the LSLFM system, with the equivalent FOV of a 40X light microscope denoted for context. The actual FOV captured by the LSLFM system is $\sim$ 140 times larger than the FOV of a 40X light microscope.  As such, the proposed LSLFM system can not only provide the tremendous advantage of a large FOV, but also may facilitate the observation of live cells in three-dimensional structures or seeded on thick or curved surfaces which is difficult to achieve with traditional optical microscopes.
%

F.K. and A.W. conceived and designed the concept.  F.K. designed the multispectral lensfree microscopy system.  A.W. designed the fusion and reconstruction algorithms.  Y.M. constructed the pulsed light source.  C.J. and S.M. performed the sample preparation. F.K. performed the data collection.  A.W. performed the data processing.  F.K., C.J., S.M., and A.W performed the data analysis.  All authors contributed to the writing and editing of the paper. We thank Maria M. F. Mesquita for her assistance with preparation of the algae samples. This work was supported by the Natural Sciences and Engineering Research Council of Canada, Canada Research Chairs Program, and the Ontario Ministry of Research and Innovation.

\pagebreak

\end{document}